\documentclass[preprint,11pt,showkeys,showpacs,nofootinbib,floatfix]{revtex4}
\usepackage{amssymb}
\usepackage{amsfonts}
\usepackage{amsmath}
\usepackage{graphicx}
\usepackage{subfigure}
\usepackage[dvipdfm,
            pdfstartview=FitH,
            bookmarksnumbered=true,
            bookmarksopen=true,
            colorlinks,
            pdfborder=001,
            linkcolor=green,
            anchorcolor=green,
            citecolor=red
            ]{hyperref}
\usepackage[toc,page,title,titletoc,header]{appendix}
\begin{document}
\title{ Holographic $s$-wave superconductors with conformal anomaly correction}
\author{Jun-Wang Lu$^{1}$}
\thanks{E-mail address:lujunwang.2008@163.com}
\author{Huai-Fan Li$^{2}$}
\thanks{E-mail address:huaifan.li@stu.xjtu.edu.cn}
\author{Ya-Bo Wu$^{3}$}
\thanks{E-mail address:ybwu61@163.com}

\affiliation{$^{1}$School of Physics and Electronics, Qiannan Normal University for Nationalities, Duyun 558000, PR~China\\
$^{2}$Department of Physics, Shanxi Datong University,  Datong 037009, PR~China\\
$^{3}$Department of Physics, Liaoning Normal University, Dalian 116029, PR~China}
\begin{abstract}
We build a holographic $s$-wave conductor/superconductor model and an insulator/superconductor model in the four-dimensional conformal anomaly corrected~(CAC) AdS gravity. The effects of CAC parameter $\alpha$ are studied using both numerical and analytical methods in the probe approximation.
Concretely, when the CAC parameter increases, the critical temperature increases for the conductor/superconductor phase transition, while the critical chemical potential decreases for the insulator/superconductor case, which suggests that the increasing CAC parameter enhances both superconducting phase transitions. Meanwhile, below the critical temperature or beyond the critical chemical potential, the scalar hair begins to condense, and the condensed phases are found to be thermodynamically stable. The critical behaviors obtained from numerics are confirmed by our analytical analysis. For the parameters we are considering, the energy gap in the conductor/superconductor model decreases monotonically by increasing the CAC parameter, while for the insulator/superconductor model the energy of quasiparticle excitations decreases with the CAC parameter.

\end{abstract}
\maketitle
\section{Introduction}\label{introduction}
The AdS/CFT correspondence provides a powerful tool to study $d$-dimensional strongly coupled systems from its $(d+1)$-dimensional gravitational theory in AdS spacetime~\cite{Maldacena1998}. In the recent years, the correspondence and its generalized version~(the so-called gauge/gravity duality)  have been widely applied in various strongly correlated systems~\cite{Liu:2018crr,Cai:2015cya,Hartnoll:2016apf,Landsteiner:2019kxb}, especially the high temperature superconductor~\cite{Hartnoll2008,Horowitz126008}.

The high temperature superconductor~(with the critical temperature usually larger than $39K$) is  believed to involve strong interaction. To understand its properties and, in particular, the microscopic mechanism are still one of the biggest challenges in condensed matter physics. The AdS/CFT correspondence naturally opens a new window to study the properties of high temperature superconductors.
By using the Einstein-Abelian-Higgs system,  the authors of Refs.~\cite{Hartnoll2008,Hartnoll:2008kx} built a holographic description of the $s$-wave conductor/superconductor model, where the main characters of a superconductor were reproduced, such as the appearance of the condensate accompanied by spontaneously breaking of the $U(1)$ symmetry below the critical temperature and the infinite DC conductivity in the broken phase. Soon after, the Meissner effect in the presence of a background magnetic field was observed via the holographic setup~\cite{Maeda:2009vf}, following which the numerical results about holographic superconductor models were confirmed by the analytical Sturm-Liouville~(S-L) eigenvalue method~\cite{Siopsis:2010uq}.

The above studies showed that the main properties of superconductors have been disclosed successfully by the gauge/gravity duality. Since then, holographic superconductors were studied from various perspectives in the literature, and an interesting direction is to construct holographic models that are more close to the real superconducting materials in condensed matter. Following this idea, the superconductor models were generalized to the SU(2) $p$-wave model~\cite{Gubser:2008wv}, the $d$-wave model~\cite{Chen:2010mk}, Maxwell-complex-vector~(MCV) $p$-wave model~\cite{Cai:2013pda,Cai:2013aca}, the superfluid model~\cite{Herzog:2008he,Basu:2008st,Wu:2014cfa,Xia:2019eje}, the coexistence and competition of multiple order parameters~\cite{Cai:2013wma,Kiritsis:2015hoa,Nie:2014qma,Nie:2020lop}, superconductor models with spatial modulations~\cite{Ling:2014laa,Cai:2017qdz,Ling:2019gjy,Cremonini:2016rbd,Cremonini:2019fzz}, the insulator/superconductor model~\cite{Nishioka131,Cai:2011ky,Li:2013fza}, complexity and entanglement entropy of superconductors~\cite{Yang:2019gce,Cai:2012sk,Cai:2012nm,Albash:2012pd}, the superconductor with the anisotropic scaling~\cite{Brynjolfsson065401,ZYFan2000,Lu:2013tza} and Kibble-Zurek Scaling~\cite{Chesler:2014gya,Zeng:2019yhi,Natsuume:2017jmu,Bu:2019epc}.

The other direction of the development is to improve the basic framework of the gauge/gravity by investigating more general interactions and corrections, in particular the terms due to
quantum corrections. The related works involve: (1) high curvature corrections such as Gauss-Bonnet gravity~\cite{Cai:2010zm,Cai:2012vk}, Quasi-topological gravity~\cite{Kuang:2010jc}, (2) nonlinear electrodynamics, for example, Born-Infeld correction~\cite{Mohammadi:2018hxc}, exponential correction, Logarithmic correction~\cite{Cheng:2018zjv}, as well as (3) gravity-gauge field coupled correction, for instance, $RF^2$~\cite{Lv:2020ecm,Lu:2018tdo} and $CF^2$~\cite{Wu:2010vr} and $C^2F^2$ corrections~\cite{Wu:2017xki,Lu:2020phn} with $R$~($C$) denoting the curvature scalar or tensor~(Weyl tensor).
It follows that the first kind of corrections (1) inhibit both the conductor/superconductor and insulator/superconductor phase transitions~\cite{Cai:2010zm,Kuang:2010jc}, while the second kind of corrections (2) inhibit the conductor/superconductor phase transition, but do not affect the critical value of insulator/superconductor phase transition~\cite{Mohammadi:2018hxc,Cheng:2018zjv}. Meanwhile, the third kind of corrections (3) enhance the conductor/superconductor phase transition, but do not influence the $s$-wave and MCV $p$-wave insulator/superconductor phase transitions~\cite{Lv:2020ecm,Lu:2018tdo,Wu:2010vr,Wu:2017xki,Lu:2020phn}.

There are also many works considering simultaneously the effects of above directions, for example, the superconductor model with backreaction in Gauss-Bonnet gravity~\cite{Cai:2010zm}, the superconductor with momentum relaxation and Weyl correction~\cite{Ling:2016lis} and the  $p$-wave model with Weyl correction or $RF^2$ correction~\cite{Lu:2018tdo} and the $p$-wave superfluid in AdS soliton~\cite{Lv:2020ecm} .
In fact, besides corrections from  Gauss-Bonnet gravity, Quasi-topological gravity, the conformal anomaly is also a kind of interesting correction worthy to be studied~\cite{Duff:1993wm}. The usual conformal anomaly characterizes the non-vanishing trace of the effective energy-momentum tensor of conformal field theory from one loop quantum corrections~\cite{Duff:1993wm}. The conformal anomaly correction not only plays an important role in the quantum field theory in curved spaces, but also has the meaningful effects in cosmology, black hole physics, string theory and statistical mechanics~\cite{Duff:1993wm}. For example, it was argued that conformal anomaly correction  might have closed relation to the Hawking radiation of black hole in the two-dimensional spacetime~\cite{ChristensenPRD1977} and also drive the inflation in cosmology~\cite{Hawking:2000bb,Nojiri:2000gb}. As a first step to study the backreaction of the trace anomaly, the authors in Ref.~\cite{Cai:2009ua} obtained some exactly nontrivial black hole solutions to the Einstein equations with conformal anomaly correction  for the first time and found there exists a logarithmic correction to the Bekenstein-Hawking entropy. Subsequently, considering the fact that the thermodynamical properties of AdS black holes crucially depend on horizon structure, the author of Ref.~\cite{Cai:2014jea} generalized the previous black holes in Ref.~\cite{Cai:2009ua} to the case with an arbitrary  constant curvature horizon. Among these black hole solutions, the planer black hole with Ricci flat horizon provides a natural gravitational background for the holographic superconductor model.
Recently, the authors of~\cite{Glavan:2019inb} proposed a novel four-dimensional Gauss-Bonnet theory as a limiting case of the original D-dimensional theory with singular Gauss-Bonnet coupling constant. While some issues and inconsistency are still debated~\cite{Gurses:2020ofy,Mahapatra:2020rds,Shu:2020cjw,Hennigar:2020lsl,Arrechea:2020evj}, such Gauss-Bonnet gravity in four dimensional spacetime has been explored extensively in the literature.  Interestingly, the same solutions of Ref.~\cite{Cai:2014jea} have been found in the four-dimensional Gauss-Bonnet theory. Therefore, it is meaningful to ask how the curvature correction affects the holographic superconductors in the novel four-dimensional Gauss-Bonnet theory. Motivated by the above mentioned, we will construct the $s$-wave superconductor model with conformal anomaly correction in this paper  and study the CAC effects on the superconductor phase transitions. The results show that the increasing CAC parameter hinders both the conductor/superconductor phase transition in the black hole and the insulator/superconductor phase transition in the soliton.

This paper is organized  as follows. In Sec.~II, we study the $s$-wave conductor/superconductor phase transition and  calculate the optical conductivity. We also investigate the critical temperature and the critical behavior by the analytical S-L method. In Sec.~III, by constructing numerically the $s$-wave insulator/superconductor model in the CAC soliton background, we study the corresponding superconductor model by the S-L method. The final section is devoted to the conclusions and  discussions. It should be noted that while this paper was being completed, the holographic superconductors in four-dimensional Einstein-Gauss-Bonnet gravity appeared in arXiv~\cite{Qiao:2020hkx}. In our present work, the numerical part about the $s$-wave conductor/superconductor model has some overlap with Ref.~\cite{Qiao:2020hkx}.

\section{Conductor/superconductor phase transition}\label{numericalcs}
In this section, we firstly construct the holographic $s$-wave conductor/superconductor phase transition in the four-dimensional CAC black hole with the Maxwell complex scalar field via the numerical method. To verify that below the critical temperature the state with scalar hair is indeed stable, we define the grand potential density of the system and then compare the grand potential of the hairy state with the one of no hairy state, following which the frequency dependent conductivity is studied in detail. In order to check the numerical results of the model, we restudy the holographic superconductor model by the analytical S-L method.

First of all, we give the four-dimensional CAC planer black hole as~\cite{Cai:2014jea}
\begin{eqnarray}\label{CACBH}
ds^2&=&-f(r)dt^2+\frac{dr^2}{f(r)}+r^2(dx^2+dy^2),\\
 &&f(r)=-\frac{r^2}{4\alpha}\left(1-\sqrt{1+\frac{8\alpha}{l^2}\left(1-\frac{r_+^3}{r^3}\right)}\right)\nonumber,
\end{eqnarray}
where $r_+$ represents  the horizon of the black hole satisfying $f(r_+)=0$, $l^{-2}$ is related to the cosmological constant $\Lambda=-\frac{3}{l^2}$. The  effective radius of the AdS spacetime is  $L_{eff}^2=4 \alpha/(\sqrt{1+8\alpha/l^2}-1)$. Meanwhile, the CAC parameter $\alpha$ can be concretely represented as $\alpha=8\pi G \vartheta$, where $\vartheta$ is a positive constant related to the degrees of freedom of quantum fields. In the present work, we will focus on the CAC parameter space as $0.0001\leq\alpha\leq10$.
Obviously, the CAC black hole returns to the standard planer AdS black hole in the case of $\alpha\rightarrow 0$~\cite{Hartnoll2008,Horowitz126008,Siopsis:2010uq,Cai:2013pda}.

Subsequently, we take the Lagrangian density consisting of a Maxwell field and a complex scalar field~\cite{Hartnoll2008}
\begin{equation}\label{Swaveaction}
  \mathcal{L}_m=-\frac{1}{4}F_{\mu\nu}F^{\mu\nu}- D_\mu\psi(D^\mu\psi)^\ast -m^2 |\psi|^2,
\end{equation}
where $F_{\mu\nu}=\nabla_\mu A_\nu-\nabla_\nu A_\mu$ is the field strength of the gauge field $A_\mu$,  $D_\mu=\nabla_\mu-iq A_\mu$, and $m$ ($q$) is the mass (charge) of the scalar field $\psi$.  To simplify the calculation, we regarded  the matter field as the probe to  the CAC black hole, where the equations of motion for both the scalar and the gauge field decouple from the Einstein field equation and the main physics of the system is believed to be grasped at the same time. Varying the action for the Lagrangian density~(\ref{Swaveaction}) with respect to the scalar field $\psi$ and the gauge field $A_\mu$, respectively, we can read off the equations of motion of scalar field and gauge field as
\begin{eqnarray}
  D_\mu D^\mu\psi-m^2\psi&=&0,\label{EOMofpsiofSwave}\\
  \nabla^\mu F_{\mu\nu}-iq(\psi^\ast D_\nu\psi-\psi (D_\nu\psi)^\ast)&=&0.\label{EOMofphiofSwave}
\end{eqnarray}
Throughout the paper, we will set $l=1$ and $q=1$ without loss of generality.

Following the works in Refs.~\cite{Hartnoll2008,Horowitz126008}, we take the complex scalar field to be real and only turn on the time component of the  Maxwell field with other components vanishing, which are
\begin{equation}\label{psiphiansatz}
\psi=\psi(r),  \ \ \  A_\mu dx^\mu=\phi(r) dt.
\end{equation}
Combining the ansatz (\ref{psiphiansatz}) with the black hole background~(\ref{CACBH}), the concrete equations of motion in term of $\psi(r)$ and $\phi(r)$ read~\cite{Hartnoll2008,Horowitz126008}
\begin{eqnarray}
 \psi''(r)+\left(\frac{f^\prime(r)}{f(r)}+\frac{2}{r}\right)\psi'(r)+\left(\frac{\phi^2(r)}{f^2(r)}-\frac{m^2}{f(r)}\right)\psi(r)&=&0,\label{snumpsi}  \\
 \phi''(r)+\frac{2}{r}\phi'(r)-\frac{2\psi(r)^2}{f(r)}\phi(r) &=& 0,\label{snumphi}
\end{eqnarray}
where the prime denotes the derivative with respect to $r$.

 To solve the above coupled differential equations, we usually impose the boundary conditions. At the horizon $r=r_+$, we require $\phi(r_+)=0$ to satisfy the finite norm of $A_\mu$, while $\psi(r_+)$ needs to be regular.
At the infinite boundary ($r\rightarrow \infty$), $\psi(r)$ and $\phi(r)$ behave as
\begin{eqnarray}
\psi(r)&=&\frac{\psi_1}{r^{\Delta_-}}+\frac{\psi_2}{r^{\Delta_+}}+\cdots, \label{asySwaveBHSpsi}\\
\phi(r)&=&\mu-\frac{\rho}{ r}+\cdots,  \label{asySwaveBHphi}
\end{eqnarray}
 where $\Delta_\pm =\frac{1}{2}\left(3\pm\sqrt{9+4m^2L^2_{eff}}\right)$, and the constants $\psi_1$~($\psi_2$) and $\mu$~($\rho$) are interpreted as the source (the vacuum expectation value) of the dual operator $\hat{\mathcal{O}}$ and the chemical potential~(the charge density) of dual field theory, respectively. By requiring that the U(1) symmetry is broken spontaneously, we impose the source-free condition $\psi_1=0$. We will focus on $\Delta_+=\Delta=2$ throughout the paper, which implies that the mass squared of the scalar field has a relation to effective radius of the AdS spacetime $m^2L^2_{eff}=-2 $.  For the above coupled equations and the asymptotical behaviors of $\psi(r)$ and $\phi(r)$, there exists an important scaling symmetry, such as $(r,T,\mu)\rightarrow \xi (r,T,\mu),\psi_2\rightarrow \xi^{\Delta} \psi_2,\rho\rightarrow \xi^{2} \rho$ with $\xi$ a positive constant, by using which we can fix  the chemical potential $\mu$ and thus work in the grand canonical ensemble.
\subsection{Numerical part}
\begin{figure}
\begin{minipage}[!htb]{0.45\linewidth}
\centering
\includegraphics[width=2.8in]{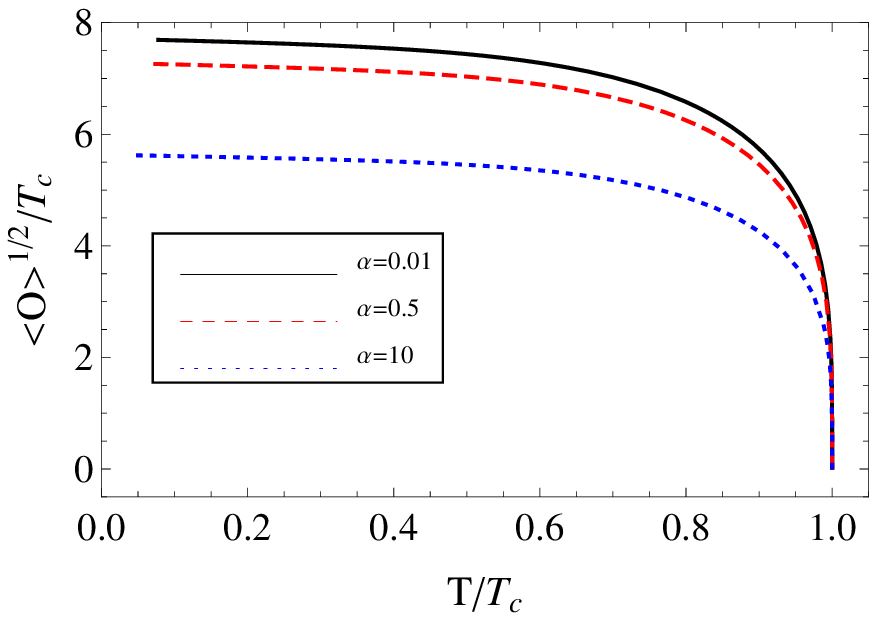}
\end{minipage}
\begin{minipage}[!htb]{0.45\linewidth}
\centering
 \includegraphics[width=2.8 in]{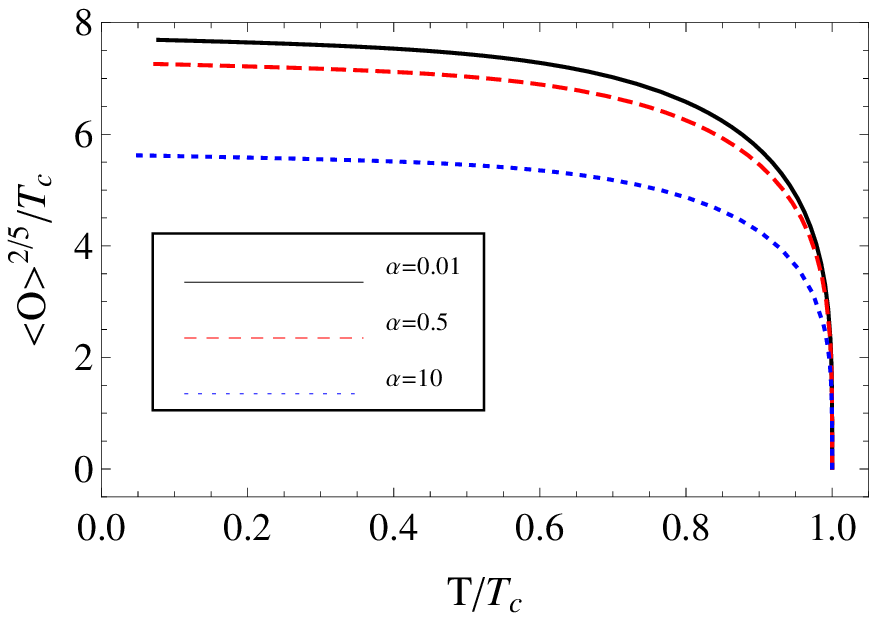}\\
\end{minipage}
\caption{The condensate  versus the temperature with $\alpha=0.01$~(black solid), $\alpha=0.5$~(red dashed), $\alpha=10$~(blue dotdashed) for $\Delta=2$~(the left panel) and $\Delta=\frac{5}{2}$~(the right panel). }
\label{condoffixDelta}
\end{figure}

\begin{figure}
\begin{minipage}[!htb]{0.45\linewidth}
\centering
\includegraphics[width=2.8in]{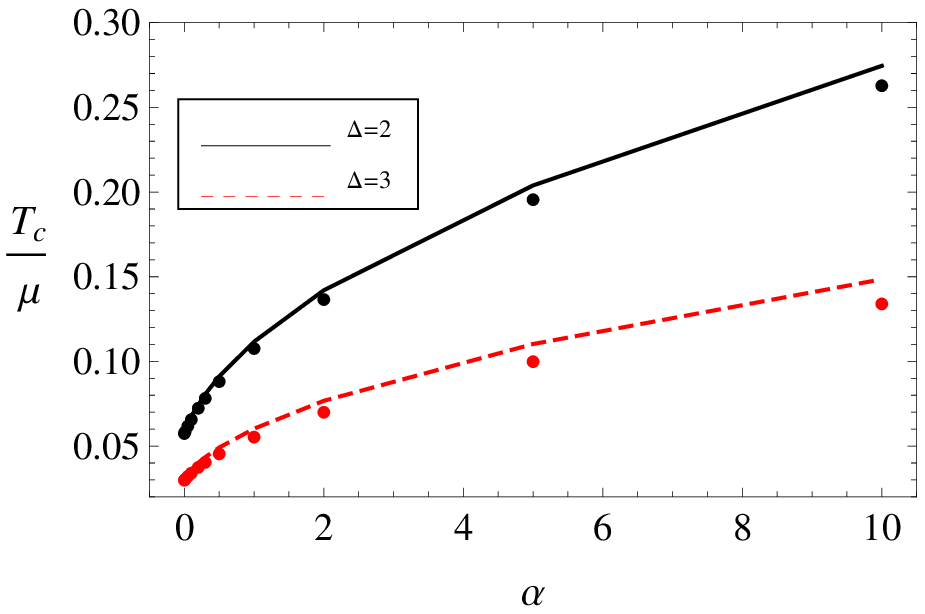}
\end{minipage}
\begin{minipage}[!htb]{0.45\linewidth}
\centering
 \includegraphics[width=2.8 in]{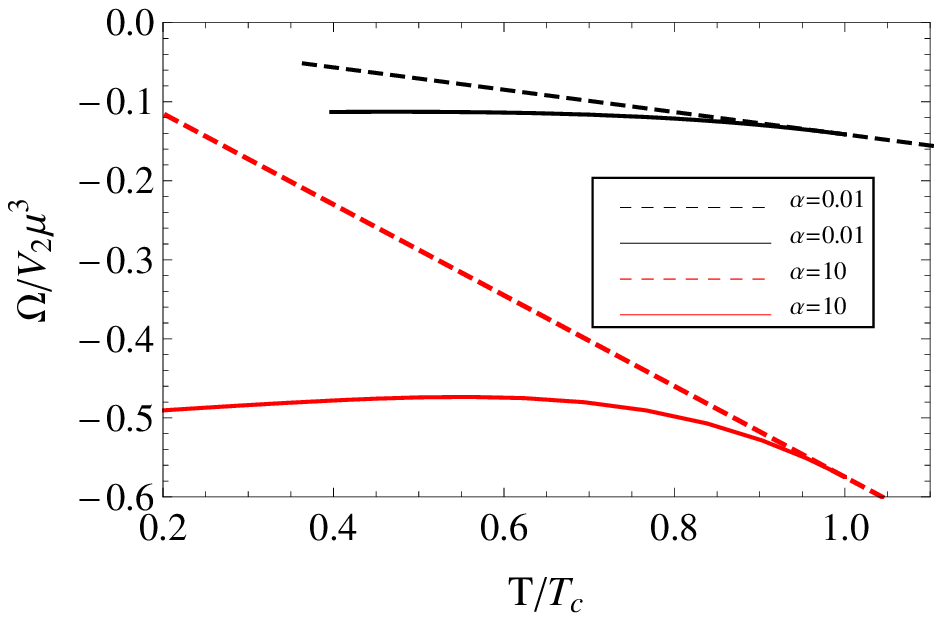}\\
\end{minipage}
\caption{The left panel represents the critical temperature  versus the CAC parameter $\alpha$ for $\Delta=2$~(black solid) and $\Delta=3$~(red dashed), while the right panel represents the grand potential about the normal state(dashed) and the superconducting state(solid) in the case of $\alpha=0.01$(black) and $\alpha=10$(red) with $\Delta=2$. }
\label{Tcandgrandpotential}
\end{figure}

After numerical calculations, we obtain the condensate with respect to the temperature for various CAC parameter $\alpha$ and scaling dimension  parameter $\Delta$. To see clearly the effect of the CAC parameter $\alpha$ on the scalar condensate,  we typically display the condensate for $\alpha=0.01,0.5,10$ with $\Delta=2$~(left panel) and $\frac{5}{2}$~(right panel) in Fig.~\ref{condoffixDelta}.
It is observed that there always exists a critical temperature below which the scalar hair starts to condense.
To study synthetically the CAC effects on the critical temperature, we list the critical temperature for various value of $\alpha$ with $\Delta=2$ in Tab.~\ref{tab:sTcna},
\begin{table}
\caption{\label{tab:sTcna} The results to different $\alpha$ for the conductor/superconductor model with $\Delta=2$: the critical temperature$\frac{T_{cn}}{\mu}$~($\frac{T_{ca}}{\mu}$) from the numerical~(analytical) method, the stable value of  $\frac{\langle \mathcal{O} \rangle^{1/2}}{T_c}$ at $\frac{T}{T_c}\approx 0.1$, the coefficient~$\mathcal{C}_{1n}$~($\mathcal{C}_{1a}$) of the condensate~$\left(\frac{\langle\hat{\mathcal{O}}\rangle}{T_c^\Delta}\approx \mathcal{C}_{1}\sqrt{1-\frac{T}{T_c}}\right)$ near the critical point from numerical (analytical)  method and the energy gap $\omega_g$.}
\begin{ruledtabular}
\begin{tabular}{c c c c c c c c }
  $\alpha$ & 0.0001 &  0.01 &0.1 & 0.3&0.5&2 &10 \\ \hline
  $\frac{T_{cn}}{\mu}$&0.05876&$0.05973$&0.06747&0.08066&$0.09110$&$0.14201$&0.27452\\
   $\frac{T_{ca}}{\mu}$&0.05739&$0.05831$&0.06562&0.07813&$0.08807$&$0.1365$&0.2628\\
   $\frac{\langle \mathcal{O} \rangle^{1/2}}{T_c}(\frac{T}{T_c}\approx 0.1)$&7.87&$7.84$&7.70&7.485&$7.244$&$6.563$&5.636\\
   $\mathcal{C}_{1n}(T\approx T_c)$&114.921 &114.700 &112.053 &106.263 &101.712 &84.496 &62.160 \\
   $\mathcal{C}_{1a}(T\approx T_c)$&59.596&59.401&57.612&54.312&51.857&42.861&31.413\\
   $\frac{\omega_g}{T_c}(\frac{T}{T_c}\approx 0.1)$&8.87&$8.80$&7.80&6.50&$5.70$&$3.65$&1.85\\
\end{tabular}
\end{ruledtabular}
\end{table}
and also show the  critical temperature versus the CAC  parameter $\alpha$ for different value of $\Delta$ in the left panel of Fig.~\ref{Tcandgrandpotential}. It is clear that the critical temperature increases with the increasing CAC parameter, which means the increasing CAC parameter enhances the conductor/superconductor phase transition. While for the fixed CAC parameter, we find that the critical temperature decreases with the increasing $\Delta$, which means that the lager mass squared of the scalar field makes the conductor/superconductor phase transition more difficult.
Meanwhile, by fitting the condensate curve near the critical point,  we find all curves of condensate versus the temperature have a square root behavior near the critical value, i.e., $\frac{\langle\hat{\mathcal{O}}\rangle}{T_c^\Delta}\approx \mathcal{C}_{1n}\sqrt{1-\frac{T}{T_c}}$,  which indicates that the system might suffer from a second-order phase transition at the critical temperature. What is more, we list the coefficient $\mathcal{C}_{1n}$ with $\Delta=2$ in Tab.~\ref{tab:sTcna}, which decreases with the increasing CAC parameter and then implies the condensate increases more and more slowly with the larger CAC parameter. Especially,  as $\alpha=0.0001$,  if we fix the charge density but not the chemical potential, we can obtain the critical temperature $\frac{T_c}{\sqrt{\rho}}=0.1184$, which obviously returns to the standard AdS black hole~\cite{Hartnoll2008,Horowitz126008,Lu:2013tza}.
Furthermore, at the lower temperature region, the scalar condensate approximates a stable value listed in Tab.~\ref{tab:sTcna}, which decreases with the increasing CAC parameter $\alpha$ and is consistent with the behavior of the condensate at the intermediate  temperature region. If we regard the value of condensate at low temperature as the condensed gap, we can see that the larger the CAC parameter, the smaller the condensed gap.
In addition, we also consider the case for other value of $\alpha$ and $\Delta$, the results show that the effects of the CAC parameter on the condensate is qualitative the same. For example, the curve of the critical temperature versus the CAC parameter with $\Delta=\frac{5}{2}$  lies between the curves of $\Delta=2$ and $\Delta=3$ in the left panel of Fig.~\ref{Tcandgrandpotential}. However,
it should be noted that when the scaling dimension parameter $\Delta$ is large enough~(for instance, $\Delta=3$), the numerical calculation becomes obviously difficult, especially for the lower temperature region of the model.

To check that below the critical point the superconducting state is indeed thermodynamically favored compared with the normal state, it is helpful to calculate the grand potential of the system, which is defined by the Euclidean on-shell action $S_E$ timing the temperature of the black hole, i.e., $\Omega=T S_E$. Integrating the Minkowski action~(\ref{Swaveaction}) by parts yields the on-shell part of action as
\begin{eqnarray}
 S_{os}&=&\int \sqrt{-g}d^4x\Big{(}-\frac{1}{2} \nabla_\mu(A_\nu F^{\mu\nu})-\nabla_\mu(\psi^\ast D^\mu\psi) +q^2\psi^2A_\nu A^\nu\Big{)}\nonumber \\
  &=&\frac{V_2}{T} \sqrt{h}n_r\left(-\frac{1}{2}A_\nu  F^{r\nu}-\psi^\ast D^r\psi\right)\Big{|}_{r\rightarrow\infty}
  +q^2\frac{V_2}{T}\int^\infty_{r_+}\sqrt{-g}\psi^2\phi^2 dr,\nonumber
\end{eqnarray}
where we have taken into account $\int d^3x=\frac{V_2}{T}$ and also Eqs.~(\ref{EOMofpsiofSwave}) and (\ref{EOMofphiofSwave}). Keepping in mind that  $S_E=-S_{os}$, we obtain the density of the grand potential as
\begin{equation}\label{grandpotential}
\frac{\Omega}{V_2}=\frac{-TS_{os}}{V_2}=-\frac{1}{2}\mu\rho+\int ^\infty_{r_+}\frac{r^2\psi^2\phi^2}{f}dr.
\end{equation}
We typically display the grand potential as a function of the temperature for the case of $\alpha=0.01$(black) and $\alpha=10$(red) with $\Delta=2$ in the right panel of Fig.~\ref{Tcandgrandpotential}. It follows that near the critical temperature, both solid curves corresponding to the superconducting state stretch out from the dashed curves corresponding to the normal state smoothly with the decreasing temperature, which means that at the critical temperature the system indeed suffers from a second-order phase transition, and thus agrees with the behavior of the condensate in Fig.~\ref{condoffixDelta}. Most importantly, the value of the grand potential of the superconducting state  is always  lower than that of the normal state, which means that below the critical temperature, the superconducting state is indeed thermodynamically stable. In addition, we also consider the other parameter cases, such as $(\Delta=2,\alpha=1,2,5,8)$ and $(\Delta=\frac{5}{2},\alpha=0.01,1,2,5,8,10)$, the curves are similar to the ones in Fig.~\ref{Tcandgrandpotential}. As a result, it is believed our numerical results are reliable in the parameter space($0.0001\leq \alpha \leq 10$).

\begin{figure}
\begin{minipage}[!htb]{0.45\linewidth}
\centering
\includegraphics[width=2.8in]{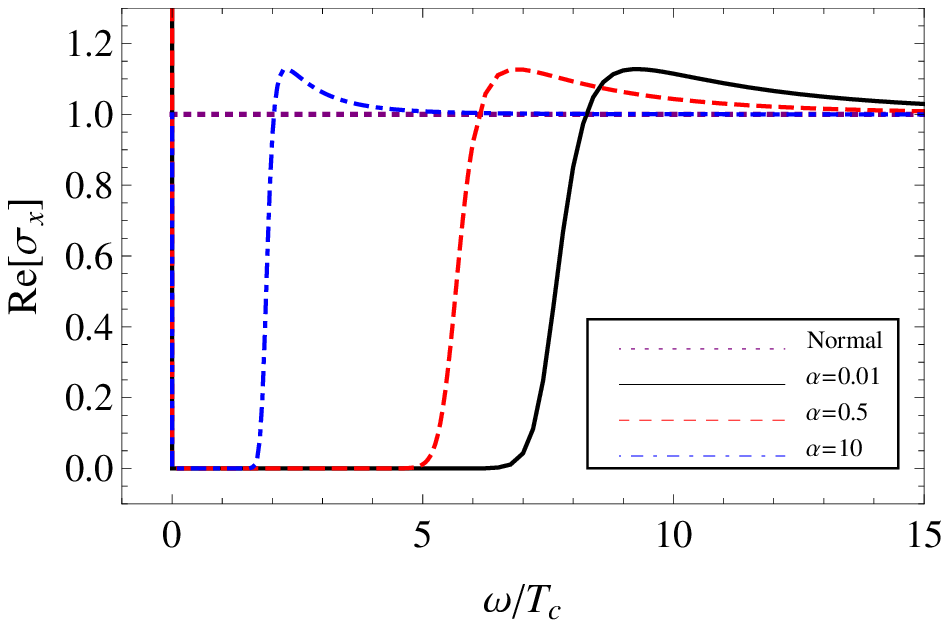}
\end{minipage}
\begin{minipage}[!htb]{0.45\linewidth}
\centering
 \includegraphics[width=2.8 in]{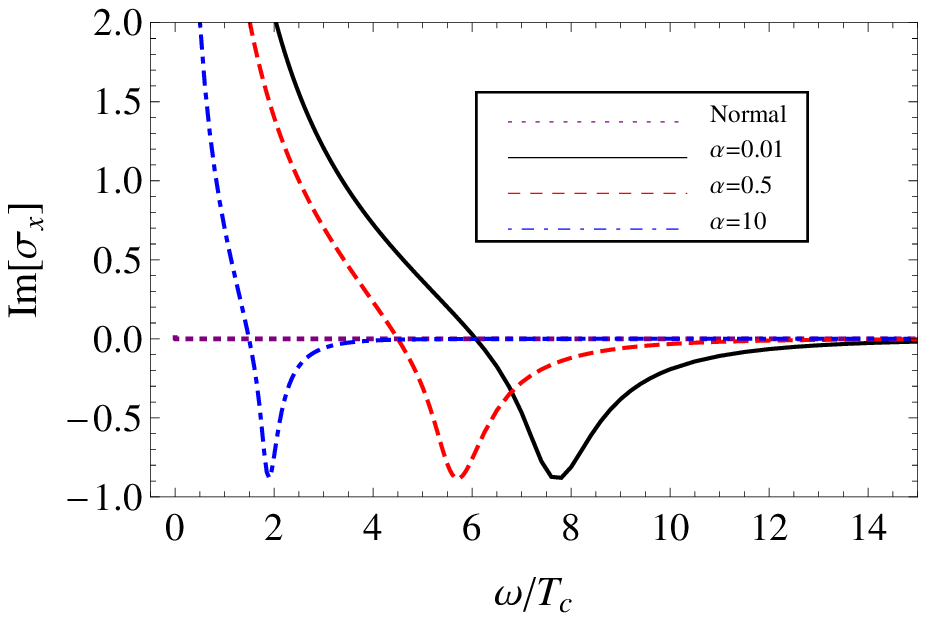}\\
\end{minipage}
\caption{The real part(left) and imaginal part(right) of the AC conductivity at $\frac{T}{T_c}\approx 0.1$ as a function of the frequency for fixed $\Delta=2$ with $\alpha=0.01$ (black), $\alpha=0.5$ (red),$\alpha=10$ (blue). }
\label{conducfixgamma}
\end{figure}

To testify the superconducting signal characterized by the infinite DC conductivity  and   the strength of the interaction in the superconductor represented by the energy gap, it is useful to compute the AC conductivity of the superconductor model. From the AdS/CFT correspondence, we study the perturbation of the gauge field in the bulk. For simplicity, we turn on the perturbation along the $x $ direction with the ansatz $\delta A(t,r)=A_x(r)e^{-i \omega t}dx$.
 The linearized equation of the perturbation $A_x(r)$  reads
\begin{eqnarray}\label{Eomax}
 &&A_x''(r)+\frac{f^\prime(r)}{f(r)} A_x'(r)+ \left(\frac{\omega^2}{f^2(r)}-\frac{2\psi^2(r)}{f(r)}\right)A_x(r)=0.
\end{eqnarray}
At the horizon, we impose the ingoing  wave condition
\begin{equation}
A_x(r)=(r-r_+)^{-i\omega/{3r_+}}\left(1+A_{x1} (r-r_+)+A_{x2} (r-r_+)^2+A_{x3}(r-r_+)^3+\cdots\right).
\end{equation}
While at the boundary, the asymptotical solution of $A_x(r)$ is given by
\begin{equation}\label{Ayasyr}
A_x(r)=A^{(0)}+\frac{ A^{(1)}}{r}+\cdots,
  \end{equation}
where $A^{(i)}$ are all constants.
Combining with Eqs.~(\ref{EOMofphiofSwave}) and ~(\ref{Ayasyr}), we can obtain the retarded Green's function as
\begin{equation}\label{eq39}
G(\omega)=-f(r)\frac{A_x^\prime(r)}{A_x(r)}\Big{|}_{r\rightarrow\infty}=\frac{1}{L_{eff}^2}\frac{A^{(1)}}{A^{(0)}}.
  \end{equation}
According to the Kubo formula, the AC conductivity reads
\begin{eqnarray}\label{conductivityformula}
\sigma(\omega)&=&-\frac{i}{\omega}G(\omega)=-\frac{i}{\omega L_{eff}^2} \frac{A^{(1)}}{A^{(0)}}.
\end{eqnarray}
 In  Fig.~\ref{conducfixgamma}, we plot the AC conductivity at $\frac{T}{T_c}\approx \frac{1}{10}$ for  $\Delta=2$ with $\alpha=0.01$~(black), $\alpha=0.5$~(red), $\alpha=10$~(blue).
It is observed from the imaginal part of conductivity at the lower frequency  there is an obvious pole for all values of the CAC parameter $\alpha$, which corresponds to a delta function of the DC conductivity as expected from  condensed physics.
Meanwhile, the imaginal part of conductivity displays a minimum at some special frequency where the real part of conductivity grows most rapidly with the increasing frequency. Following Refs.~\cite{Hartnoll2008,Horowitz126008}, we interpret the value of this special frequency as the energy gap which is believed to characterize the strength of the interaction in the superconductor. What is more, we find for fixed $\Delta=2$, the ratio of the energy gap to the critical temperature decreases with the increasing CAC parameter, which means that the larger CAC parameter suppresses the energy gap and obviously enhances the conductor/superconductor phase transition, which is consistent with the CAC effect  on the critical temperature in Fig.~\ref{Tcandgrandpotential}.  At the same time, we also list the value of the energy gap in Tab.~\ref{condoffixDelta}. It follows  that the energy gap decreases from $8.87$ to $1.85$, which seems to suggest that the present superconductor model not only displays the high-temperature superconductor with strong interaction but also the conventional BCS superconductor.
Moreover, at the high frequency region, the conductivity extends a stable value, which is the universal behavior in the four dimensional black hole case. In addition, we also consider the case with different CAC parameter, and obtain  the similar behavior  in Fig.~\ref{conducfixgamma}.

\subsection{Analytical part}
By means of the new variable $z=\frac{r_+}{r}$, Eqs.~(\ref{snumpsi}) and (\ref{snumphi}) can be rewritten  as
 \begin{eqnarray}
 \psi''(z)+\frac{f'(z)}{f(z)}\psi'(z)+\left(\frac{\phi(z)^2}{r_+^2z^4f^2(z)}-\frac{m^2}{ z^4f(z)}\right)\psi(z)&=&0,\label{BHpsiz}  \\
 \phi''(z)-\frac{2 \psi (z)^2}{z^4f(z)}\phi(z) &=& 0,\label{BHphiz}
\end{eqnarray}
where the prime represents the derivative with respect to the variable $z$.

At the critical point, the scalar condensate vanishes, we can thus read off the solution of the gauge field as
\begin{equation}\label{phianala}
\phi(z)=\lambda r_+(1-z), \lambda=\frac{\mu}{r_{+c}},
\end{equation}
where $r_{+c}$ represents the location of the horizon at the critical temperature.

Near the critical point,  we can express the scalar field $\psi(z)$ as
\begin{equation}\label{psianaans}
\psi(z)= \frac{\langle \mathcal{O}\rangle}{r_+^\Delta} z^\Delta F(z),
\end{equation}
with the boundary condition $F(0)=1$ and $F'(0)=0$~\cite{Siopsis:2010uq,Li:2013fza,ZYFan2000,Lu:2013tza}. Taking into account Eq.~(\ref{phianala}) and substituting Eq.~(\ref{psianaans}) into Eq.~(\ref{BHpsiz}) yields the equation in term of $F(z)$ as
\begin{equation}\label{anacsbhF}
F''+\left(\frac{2\Delta}{z}+\frac{f'(z)}{f(z)}\right)F'+\left(\frac{(1-z)^2}{z^4f^2(z)}\lambda^2+\frac{\Delta z^2(zf'(z)+(\Delta-1)f(z))-m^2}{z^4f(z)}\right)F=0,
\end{equation}

By multiplying $\mathcal{T}= z^{2 \Delta}f(z)$ to Eq.~(\ref{anacsbhF}) reads the S-L eigenvalue equation as
\begin{equation} \label{SolitonSLeq}
\frac{d}{dz}(\mathcal{T} F')-\mathcal{P}F +\lambda^2 \mathcal{Q}F=0,
\end{equation}
 where the coefficients $\mathcal{P}$ and $\mathcal{Q}$ are given by
\begin{equation}
\mathcal{P}=-z^{2 \Delta-4 }( \Delta z^2 ((\Delta-1)f(z)+zf'(z))-m^2),\ \ \  \  \mathcal{Q}= \frac{(1-z)^2 z^{2 \Delta -4}}{f(z)}.
\end{equation}
Thus $\lambda^2$ is obtained by minimizing the following expression as~\cite{Siopsis:2010uq,Lu:2013tza}
\begin{equation}\label{eigenvalueexpress}
\lambda^2=\frac{\int_0^1 (\mathcal{T} {F'}^2-\mathcal{P}F^2)dz}{\int^1_0 \mathcal{Q} F^2 dz}.
\end{equation}
Given the boundary conditions $F(0)=1,~F'(0)=0$,
 we can take the form of trial function as
 \begin{equation} \label{trialfunction}
F=F_\beta(z)\equiv 1-\beta  z^2,
\end{equation}
by plugging which into Eq.~(\ref{eigenvalueexpress}) we can obtain  the value of $\lambda$ with respect to some special $\beta$. Thus the critical temperature can be written as
\begin{equation}
T_c=\frac{3}{4\pi}r_{+c}=\frac{3}{4\pi}\frac{\mu}{\lambda}.
\end{equation}
The concrete analytical values of the critical temperature  $\frac{T_{ca}}{\mu}$ are listed in Tab.~\ref{tab:sTcna},
from which we can see clearly that the critical temperature increases with the increasing CAC parameter $\alpha$, which means that the larger CAC parameter makes the conductor/superconductor phase transition easier to occur. Meanwhile, the analytical results agree well with the numerical ones. In particular, as $\alpha=0.0001$, the result almost restores to the standard AdS black hole case in Refs.~\cite{Hartnoll2008,Horowitz126008,Siopsis:2010uq,Lu:2013tza}.

Below (but close to) the critical temperature, the condensate $\frac{\langle\mathcal{O}\rangle}{r_+^\Delta}$ is very small,  by using which the gauge field (\ref{phianala}) can be expanded in the form of the small parameter as
\begin{equation}\label{phianacsO}
 \frac{\phi(z)}{r_+}=\lambda(1-z)+\frac{\langle\mathcal{O}\rangle}{r_+^\Delta}\chi(z).
\end{equation}
Substituting  Eq.~(\ref{phianacsO}) into  Eq.~(\ref{BHphiz}), we can obtain the equation of $\chi(z)$  at the linear order of $\frac{\langle\mathcal{O}\rangle}{r_+^\Delta}$ as
\begin{equation}\label{chi2csana}
\chi''(z)=  \lambda  \frac{\langle\mathcal{O}\rangle}{r_+^\Delta}  \frac{ 2(1-z) F^2(\beta,z) z^{2 \Delta -4}}{f(z)}
\end{equation}
with the boundary condition $\chi(1)=0=\chi'(1)$~\cite{Siopsis:2010uq,Lu:2013tza}.
Integrating Eq.~(\ref{chi2csana}) yields
\begin{equation}\label{chiuBH}
\chi'(u)=- \lambda\frac{\langle\mathcal{O}\rangle} {r_+^\Delta}  \int^{z=1}_{z=u}\frac{2(1-z) F^2(\beta,z) z^{2 \Delta -4}}{f(z)}dz,
\end{equation}
by further integrating which we have
\begin{equation}\label{anacschi0}
\chi(0)=\lambda\frac{\langle\mathcal{O}\rangle} {r_+^\Delta}  \int^{u=1}_{u=0}\int^{z=1}_{z=u}\frac{2(1-z) F^2(\beta,z) z^{2 \Delta -4}}{f(z)}dzdu\equiv \lambda\frac{\langle\mathcal{O}\rangle} {r_+^\Delta}\mathcal{C}_2(\beta,\Delta).
\end{equation}

On the other hand, expanding Eq.~(\ref{phianacsO}) at the infinite boundary~($z\rightarrow0$) and comparing the zero order of $z$ with Eq.~(\ref{asySwaveBHphi}), we have
\begin{equation}\label{anacsOchi}
\frac{\langle\mathcal{O}\rangle} {r_+^\Delta}\chi(0)=\frac{\mu}{r_+}-\lambda.
\end{equation}

Taking notice of the value of $\chi(0)$ in Eq.~(\ref{anacschi0}), the condensate near the critical point can be expressed as
\begin{equation}
\left(\frac{\langle\mathcal{O}\rangle} {r_+^\Delta}\right)^2=\frac{1}{\mathcal{C}_2(\beta,\Delta)}\frac{T_c}{T}\left(1-\frac{T}{T_c}\right)\approx \frac{1}{\mathcal{C}_2(\beta,\Delta)}\left(1-\frac{T}{T_c}\right),
\end{equation}
where we have considered the approximation $T\approx T_c$. Therefore, the critical behavior of the condensate is given by
\begin{equation}
\frac{\langle\mathcal{O}\rangle} {{T_c}^\Delta}=\left(\frac{4\pi}{3}\right)^\Delta\frac{1}{\sqrt{\mathcal{C}_2(\beta,\Delta)}}\sqrt{1-\frac{T}{T_c}}=C_{1a}\sqrt{1-\frac{T}{T_c}}.
\end{equation}
The coefficient $\mathcal{C}_{1a}$ is listed in Tab.~\ref{tab:sTcna}, from which we find that the coefficient agrees with the numerical results at the same order, especially, the monotonically decreasing trend as a function of the CAC parameter $\alpha$.

\section{Insulator/superconductor phase transition}\label{inssuper}

Similar to the idea of the conductor/superconductor model based on the CAC black hole in the above section, we will numerically build the corresponding insulator/superconductor phase transition in the four-dimensional AdS soliton and then calculate the grand potential of the system, following which we study the frequency dependent conductivity. To back up the numerical results, we will reconstruct the holographic superconductor model by the analytical S-L method.

By performing the double Wick rotation~($t\rightarrow i\eta,~y\rightarrow i t$) to the four-dimensional CAC black hole solution~(\ref{CACBH}), a four-dimensional CAC soliton is of the form~\cite{Nishioka131,Cai:2011ky,Li:2013fza,Lu:2013tza}
\begin{eqnarray}\label{AdSsolitonmetric}
ds^2&=&-r^2dt^2+\frac{dr^2}{f(r)}+r^2dx^2+f(r)d\eta^2,
f(r)=-\frac{r^2}{4\alpha}\left(1-\sqrt{1+\frac{8\alpha}{l^2}\left(1-\frac{r_0^3}{r^3}\right)}\right),
\end{eqnarray}
where $r_0$ denotes the tip of the soliton geometry to distinguish the soliton from the black hole.  To have a smooth geometry, we impose a periodicity $\eta\sim\eta+\frac{\pi}{r_0}$ for the Scherk-Schwarz circle on the spatial direction $\eta$. Due to the fact that the soliton solution has no horizon, thus no temperature exists. Meanwhile, because of the tip for the soliton geometry, there is an IR cutoff (mass gap) for the dual field theory, which means a confined phase. Therefore, the soliton gravitational background is believed to model the  insulator in condensed matter physics~\cite{Nishioka131,Cai:2011ky,Li:2013fza}. In addition, because of the compactness of the spatial direction $\eta$, the present four-dimensional soliton background is indeed dual to a two-dimensional field theory with mass gap.

Following the works in Refs.~\cite{Nishioka131,Cai:2011ky,Li:2013fza}, we take the form of the complex scalar field and the  Maxwell field the same with Eq.~(\ref{psiphiansatz}) and thus obtain
 the concrete equations of motion in term of $\psi(r)$ and $\phi(r)$ in the CAC soliton background~(\ref{AdSsolitonmetric}) as
\begin{eqnarray}
 \psi''(r)+\left(\frac{2}{r}+\frac{f'(r)}{f(r)}\right)\psi'(r)+\left(\frac{\phi^2}{r^2f(r)}-\frac{ m^2 }{ f(r)}\right)\psi(r)&=&0,\label{snumpsis}  \\
 \phi''(r)+\frac{f'(r)}{f(r)}\phi'(r)-\frac{2\psi(r)^2}{f(r)}\phi(r) &=& 0,\label{snumphis}
\end{eqnarray}
where the prime stands for the derivative with respect to $r$.
\subsection{Numerical part}\label{numericalss}
To solve numerically the coupled differential equations (\ref{snumpsis}) and (\ref{snumphis}), we have to specify the boundary conditions for $\psi(r)$ and $\phi(r)$.  It should be noted that the constant $\mu$ is a trial solution of Eq.~(\ref{snumphis}). Different from the AdS black holes requiring the gauge field to be zero at the horizon~\cite{Hartnoll2008,Horowitz126008}, here we only impose the Neumann-like boundary condition~\cite{Nishioka131} to remove the logarithm term  in order for both $\psi(r_0)$ and $\phi(r_0)$  to be finite at the tip $r=r_0$.
At the infinite boundary~($r\rightarrow \infty$), $\psi(r)$ and $\phi(r)$ have the same asymptotical expansions  as Eqs.~(\ref{asySwaveBHSpsi}) and~(\ref{asySwaveBHphi}) in the black hole case.
 According to the gauge/gravity duality, the coefficients~($\psi_1$ ($\psi_2$) and $\mu$($\rho$)) of the insulator/superconductor model have the same physical explanations with the conductor/superconductor model in Sec.~\ref{numericalcs}. Since the U(1) symmetry is expected to be broken spontaneously, we still impose the source-free condition $\psi_1=0$. Hereafter, we still concentrate on the  $\Delta=2$ case in the present section and take $r_0=1$ in the numerical calculation.
Thus the period of the spatial coordinate $\eta$ is
 $\Gamma=\pi$. In order for the insulator/superconductor model with different CAC parameter to be comparable, we fix the period of the spatial coordinate $\eta$, which is another difference from the case of the black holes requiring either the charge density or the chemical potential to be fixed. However, the period $\Gamma$  is obviously independent of the CAC parameter, which implies that we do not need rescale the numerical results if we fix $\Gamma=\pi$ in the current paper.

\begin{figure}
\begin{minipage}[!htb]{0.45\linewidth}
\centering
\includegraphics[width=3.1in]{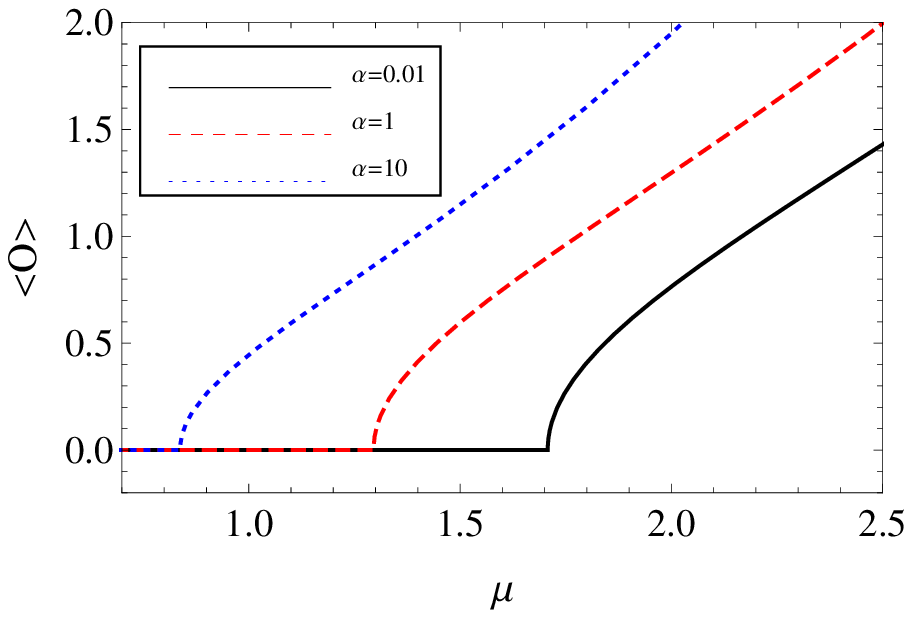}
\end{minipage}
\begin{minipage}[!htb]{0.45\linewidth}
\centering
\includegraphics[width=3.1in]{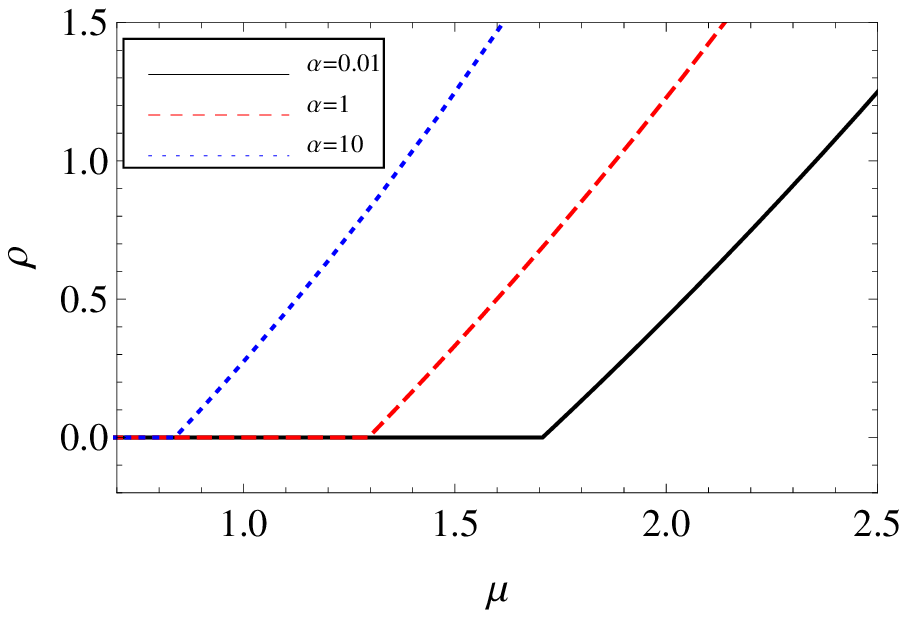}
\end{minipage}
\caption{The condensate (left) and the charge density (right) versus chemical potential with $\alpha=0.01,1,10$ (from right to left) for fixed $\Delta=2$.}
\label{condchargedensity}
\end{figure}

\begin{figure}
\begin{minipage}[!htb]{0.45\linewidth}
\centering
\includegraphics[width=2.8in]{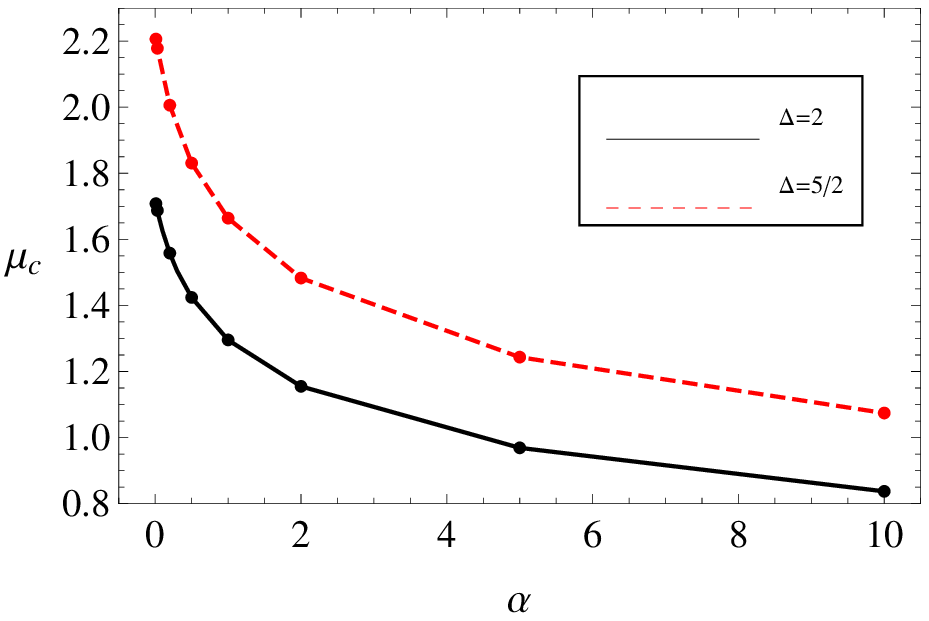}
\end{minipage}
\begin{minipage}[!htb]{0.45\linewidth}
\centering
 \includegraphics[width=2.8 in]{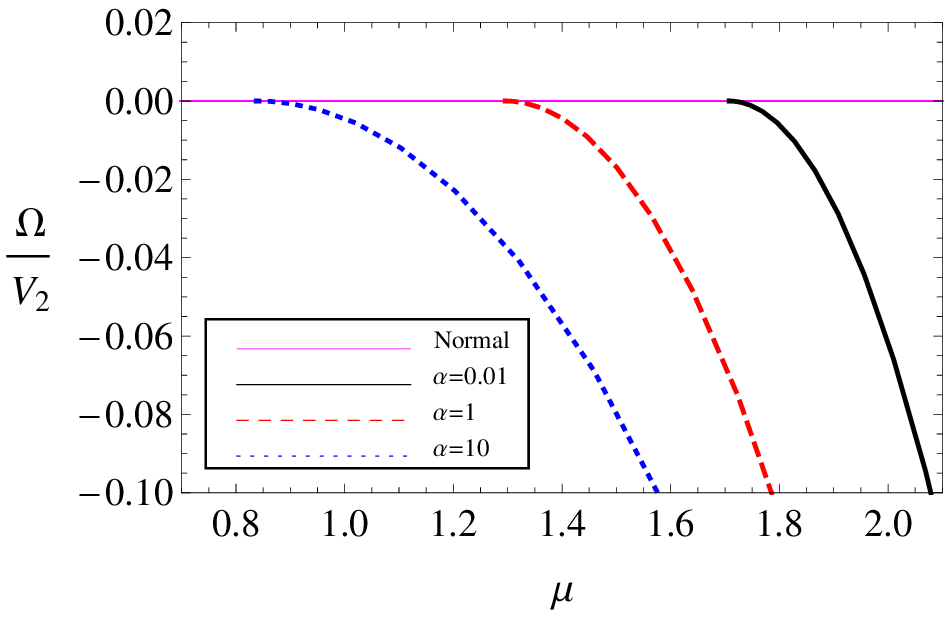}\\
\end{minipage}
\caption{The left panel represents the critical chemical potential  versus the CAC parameter $\alpha$ in the case of $\Delta=2,\frac{5}{2}$, while the right panel shows the grand potential as a function of the chemical potential about the normal state(horizontal) and the superconducting state(curved) in the case of $\alpha=0.01,1,10$(from right to left) with fixed $\Delta=2$. }
\label{Criticalmugrandpotential}
\end{figure}

We plot the scalar condensate $\langle \mathcal{O}\rangle$ and the  charge density $\rho$ versus the chemical potential with $\alpha=0.01$~(black solid),~$1$~(red dashed) and ~$10$~(blue dotted) in Fig.~\ref{condchargedensity},
from which we have the following remarks. As for the curves of  scalar condensate, first of all, there always exists a critical chemical potential for all cases, above which the scalar hair starts to condense.
 Meanwhile, we have also listed the critical chemical potential in Tab.~\ref{criticalmuc1c2}
\begin{table}
\caption{\label{criticalmuc1c2} The results for the insulator/superconductor model with $\Delta=2$: the critical chemical potential from the numerical method ($\mu_c$). $\langle \mathcal{O} \rangle=\mathcal{C}_{3n}\sqrt{\frac{\mu-\mu_c}{\mu_c}}$ and $\rho\approx \mathcal{C}_{4n}(\mu-\mu_c)$ near the critical point~$(\mu\approx \mu_c)$, $\omega_{gn}(\frac{\mu}{\mu_c}\approx 2(5))$ denotes the location of second pole for the conductivity for $\frac{\mu}{\mu_c}\approx 2$ ($5$). }
\begin{ruledtabular}
\begin{tabular}{c c c c c c c  }
  $\alpha$ & 0.01 &  0.2 &0.5 & 1&5&10  \\ \hline
  $\mu_{cn}$&1.7074&$1.5578$&1.4238&1.2953&$0.9689$&$0.8373$\\
   $\mathcal{C}_{3n}(\mu\approx \mu_c)$&1.7739&$1.7030$&1.6110&1.4390&$1.1241$&$0.9571$\\
   $\mathcal{C}_{4n}(\mu\approx \mu_c)$&1.4573&$1.541$&1.594&1.636&$1.746$&$1.812$\\
   $\omega_{gn}(\frac{\mu}{\mu_c}\approx 2)$&2.169&$1.933$&1.745&1.574&$1.163$&$1.002$\\
  $\omega_{gn}(\frac{\mu}{\mu_c}\approx 5)$&3.985&$3.529$&3.175&2.859&$2.106$&$1.813$\\
\end{tabular}
\end{ruledtabular}
\end{table}
and plotted the value of $\mu_c$ versus the CAC parameter $\alpha$ in the form the black solid curve in the left panel of Fig.~\ref{Criticalmugrandpotential}.
 It follows that the critical chemical potential decreases with the larger CAC parameter $\alpha$, which means that the larger CAC correction enhances the insulator/superconductor phase transition. What is more, for fixed CAC parameter, the larger the dimensional scaling $\Delta$, the smaller the critical chemical potential, which is reasonable, because the larger $\Delta$ means the larger mass squared $m^2$ of the scalar field, which must hinder the scalar field to condense. Meanwhile, near the critical point, we have $\langle \mathcal{O}\rangle \sim \mathcal{C}_{3n}\sqrt{\mu-\mu_c}$ by fitting the numerical curves in Fig.~\ref{condchargedensity}. The critical exponent of the condensate ($\frac{1}{2}$) indicates that the system undergoes a second-order phase transition at the critical point. Furthermore, we read off the coefficient $\mathcal{C}_{3n}$ in Tab.~\ref{criticalmuc1c2} and find it decreases with the increasing CAC parameter, which suggests that the larger CAC parameter suppresses the growth of the condensate. In term of the curves for the charge density, we observe that above the critical point, the charge density appears and increases with the chemical potential. By fitting the numerical results, we also find that the charge density has the linear dependent on the chemical potential as $\rho\sim\mathcal{C}_{4n}(\mu-\mu_c)$, which agrees with the meaning field theory. Besides, the coefficient $\mathcal{C}_{4n}$ listed in Tab.~\ref{criticalmuc1c2} increases with the increasing CAC parameter.

In order to prove that above the critical  chemical potential, the superconducting state with scalar hairy  is indeed thermodynamically favored in contrast  with the normal state, we introduce the `temperature' of the soliton as $\int dt=\frac{1}{T}$, upon which the grand potential of the system  is defined by the Euclidean on-shell action $S_E$ timing the `temperature' of the soliton, i.e., $\Omega=T S_E$. Integrating the Minkowski action~(\ref{Swaveaction}) by parts, the on-shell action is of the form
\begin{eqnarray}
 S_{os} &=&\frac{V_2}{T} \left(\frac{\mu\rho}{2L^2_{eff}}-\int^\infty_{r_0}\psi^2\phi^2dr\right),
\end{eqnarray}
where we have taken into account $\int d^2x=V_2$ and also Eqs.~(\ref{snumpsi}) and (\ref{snumphi}). Having in mind that  $S_E=-S_{os}$, the density of the grand potential is given by
\begin{equation}\label{grandpotential}
\frac{\Omega}{V_2}=\frac{-TS_{os}}{V_2}=-\frac{\mu\rho}{2L^2_{eff}}+\int ^\infty_{r_0}r\phi^2\psi^2 dr.
\end{equation}

Because of the existence of the  effective radius of AdS spacetime~$L^2_{eff}$, it is clear that the CAC parameter $\alpha$ will affect the grand potential.  Next, we typically display the grand potential with respect to the chemical potential for the case of $\Delta=2$ in the right panel of Fig.~\ref{Criticalmugrandpotential},
from which we can observe that near the critical point, the superconducting curve stretches out from the horizontal line corresponding to insulators smoothly with the increasing chemical potential, which means that the system indeed suffers from a second-order phase transition at the critical point, and thus agrees with the behavior of the condensate in Fig.~\ref{condchargedensity}.
What is more, the value of the grand potential of the superconducting state  is always  lower than the one of the normal state, which means that above the critical value, the superconducting state is indeed thermodynamically stable. In addition, the behaviors of the other values of the CAC parameter in $0.001\leq \alpha\leq 10$ are similar to the case in Fig.~\ref{Criticalmugrandpotential}.

In what follows, we calculate the electromagnetic perturbation in the hairy soliton to study the conductivity. Concretely,  we turn on the perturbation $\delta A=A_x(r)e^{-i \omega t}dx$, and thus obtain the linearized  equation~\cite{Nishioka131,Lu:2013tza}
\begin{equation}\label{Sswaveaxeq}
A_x''(r)+\frac{f'(r)}{f(r)}A_x'(r)+\left(\frac{\omega^2}{r^2f(r)}-\frac{2\psi^2(r)}{f(r)}\right)A_x(r)=0.
\end{equation}
In order for $A_x$ to be finite at the tip, we  take the ansatz of $A_x$ near the tip
\begin{equation}\label{Solitontipcond}
A_x(r)=1+A_{x1} (r-r_0)+A_{x2} (r-r_0)^2+A_{x3}(r-r_0)^3+\cdots,
\end{equation}
where $A_{x1}$, $A_{x2}$ and $A_{x3}$ are all constants and the leading term is taken to be unity due to the linearity of the equation for $A_x$.
At the infinite boundary~($r\rightarrow\infty$), the asymptotical expansion of $A_x$ is the same with Eq.~(\ref{Ayasyr}). From the gauge field perturbation we can find that the Green function is still~(\ref{eq39}). Therefore, the formula of the frequency conductivity still equates to Eq.~(\ref{conductivityformula}).

\begin{figure}
\begin{minipage}[!htb]{0.45\linewidth}
\centering
\includegraphics[width=2.9 in]{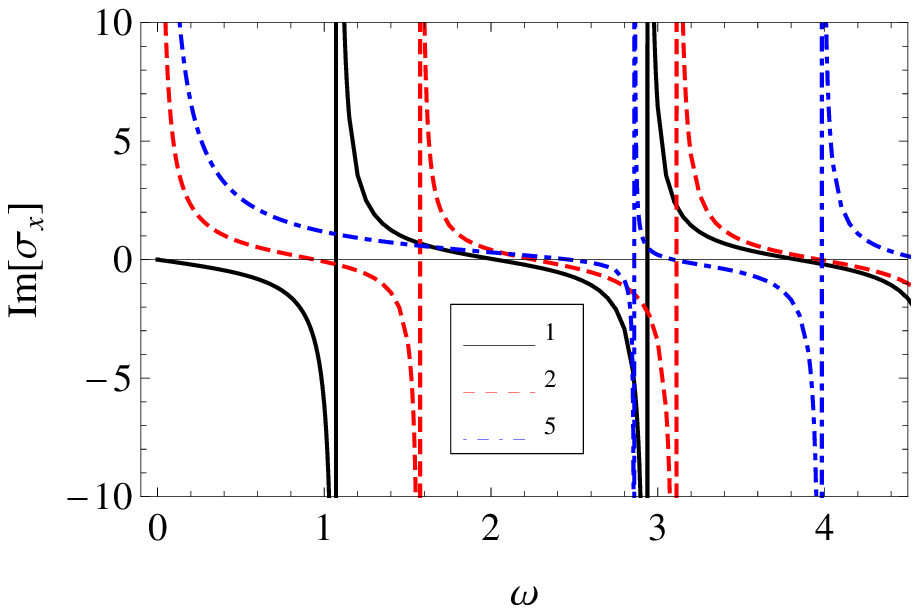}
\end{minipage}
\begin{minipage}[!htb]{0.45\linewidth}
\centering
\includegraphics[width=2.9 in]{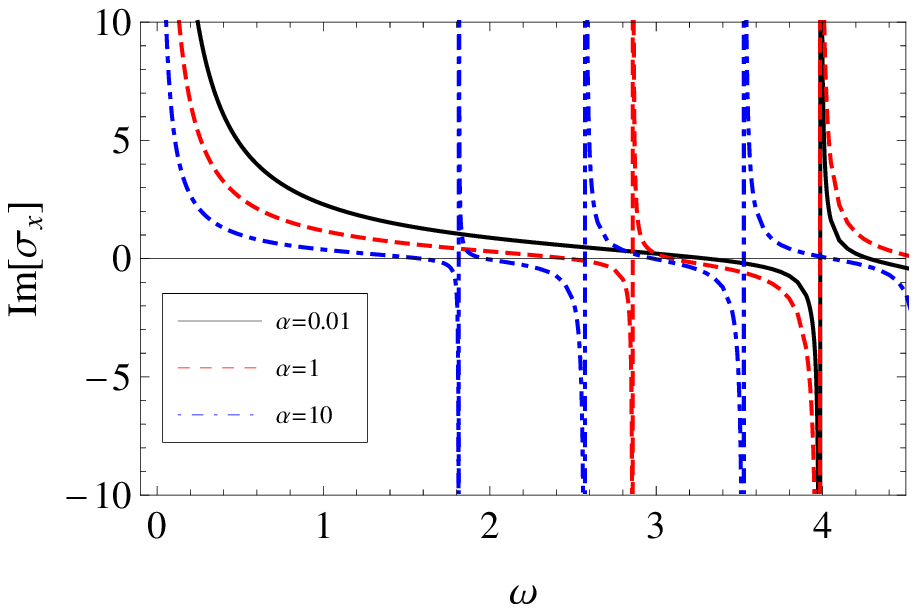}
\end{minipage}
\caption{The left panel represents the imaginal part of frequency dependent conductivity  for $\Delta=2$ and $\alpha=1$  in the case of $\frac{\mu}{\mu_c}\approx 1,~2,~5$ (from left to right), while the right panel represents the ones with  $\frac{\mu}{\mu_c}\approx 5$ and $\alpha=10,1,0.01$ (from left to right).}
\label{condufre}
\end{figure}

In Fig.~\ref{condufre}, we typically show the imaginal part of conductivity as a function of the frequency for different values of chemical potential and $\Delta$,
from which some remarks are in order. In term of the left panel, we can see that at the critical chemical potential  the imaginal part of the conductivity vanishes at the low frequency region, which corresponds to the finite  conductivity. However, when the chemical potential increases away from the critical point, such as $\frac{\mu}{\mu_c}\approx 2~(5)$, a clear pole appears in the low frequency region, which implies the infinite DC  conductivity as expected from the superconducting state. Meanwhile, the value of the location for the second pole of conductivity increases with the larger chemical potential, which suggests that the larger chemical potential increases the energy of  quasiparticle excitation.  As for the right panel, due to the fact that all curves are from superconducting state, it is reasonable that there always exists a pole in the low frequency region. Furthermore, we list the value of the second pole of the imaginal part of conductivity for various CAC parameter $\alpha$ for $\frac{\mu}{\mu_c}\approx 2~(5)$ in Tab.~\ref{criticalmuc1c2}. It follows that for the fixed ratio of the chemical potential to the critical value, the location of the second pole moves toward left when one increases the value of $\alpha$. The case of $\frac{\mu}{\mu_c}\approx 5$ has the similar behavior to the case of  $\frac{\mu}{\mu_c}\approx 2$.

\subsection{ Analytical part}
To back up the above numerical results, especially, the effects of the CAC parameter on the critical chemical potential and the condensate,  we construct the $s$-wave insulator/superconductor model by the analytical S-L method. Concretely, we resolve analytically the coupled differential equations (\ref{snumpsis}) and (\ref{snumphis}) with the same boundary conditions as the ones in subsection~\ref{numericalss}. By introducing a new variable $z=\frac{r_0}{r}$, Eqs.~(\ref{snumpsis}) and (\ref{snumphis}) can be rewritten as
\begin{eqnarray}
\psi ''(z)+\frac{f'(z)}{f(z)}\psi '(z)+\left(\frac{\phi (z)^2}{r_0^2z^2f(z)}-\frac{m^2}{ z^2f(z)}\right)\psi (z)&=&0,\label{EOMofpsiz}\\
\phi ''(z)+\left(\frac{2}{z}+\frac{f'(z)}{f(z)}\right)\phi '(z)-\frac{2  \psi ^2(z)}{z^4 f(z)} \phi (z)&=&0,\label{EOMofphiz}
\end{eqnarray}
where the prime represents the derivative with respect to $z$.

In the normal phase, $\psi(z)= 0$, the general solution $\phi(z)$ from Eq.~(\ref{EOMofphiz}) is of the form
\begin{eqnarray}
\phi(z)&=&\mathcal{C}_5 +\mathcal{C}_6\Big{(}\frac{z \left(1-z^3\right) \sqrt{8 \alpha -8 \alpha  z^3+1}
   F_1\left(\frac{1}{3};-\frac{1}{2},1;\frac{4}{3};\frac{8 z^3 \alpha }{8 \alpha +1},z^3\right)}{8 \alpha
   \left(z^3-1\right) \sqrt{1-\frac{8 \alpha  z^3}{8 \alpha +1}}}+ \nonumber\\
    &&-\frac{\log \left(z^2+z+1\right)}{48 \alpha}+\frac{\log (1-z)}{24 \alpha}-\frac{\tan ^{-1}\left(\frac{2z+1}{\sqrt{3}}\right)}{8 \sqrt{3} \alpha}\Big{)},
\end{eqnarray}
where $\mathcal{C}_5$ and $\mathcal{C}_6$ are two constants. As we have analyzed in subsection~\ref{numericalss}, we take $\mathcal{C}_6=0$  in order for the gauge field to be finite at the tip via the Neumann-like boundary condition~\cite{Nishioka131,Cai:2011ky,Li:2013fza}. Hence, the constant $\mathcal{C}_5=\mu$ is regarded as the chemical potential in the dual field theory. Obviously, the charge density vanishes in the normal phase, which agrees well with the numerical results in Fig.~\ref{condchargedensity}.

When the chemical potential goes slightly beyond the critical point, the scalar condensate begins to condense and can be expressed as
\begin{equation}\label{ConFz}
\psi= \langle \mathcal{O}\rangle z^\Delta F(z),
\end{equation}
where $F(z)$ is a function to be determined with the boundary condition $F(0)=1$. Plugging Eq.~(\ref{ConFz}) into Eq.~(\ref{EOMofpsiz}) yields the equation of $F(z)$ as
\begin{eqnarray}
F''(z)+\left(\frac{f'(z)}{f(z)}+\frac{2
   \Delta }{z}\right)F'(z)\ \ \ \ \ \ \ \ \ \ \ \ \ \ \ \ \ \ \ \ \ \ \ \ \ \ \ \ \ \ \ \ \ \ \ \ \ \ \ \ \ \ \ \ \ \ \ \  &&\nonumber \\
+\left(\frac{\Delta  z^2 \left(z f'(z)+(\Delta -1) f(z)\right)-m^2}{z^4f(z)}+\frac{\lambda ^2}{z^2 f(z)}\right)F(z)&=&0.
\end{eqnarray}
Multiplying the factor $\mathcal{T}=z^{2\Delta}f(z)$ to the above equation yields the S-L eigenvalue equation as
\begin{equation} \label{SolitonSLeq}
\frac{d}{dz}(\mathcal{T} F')-\mathcal{P}F +\mu_c^2 \mathcal{Q}F=0,
\end{equation}
where $\mathcal{P}$ and $\mathcal{Q}$ are given by
\begin{eqnarray}
\mathcal{P}&=&-z^{2 \Delta -4} \left(\Delta  z^2 \left(z
   f'(z)+(\Delta -1) f(z)\right)-m^2\right),\ \ \ \ \ \mathcal{Q}= z^{2 \Delta -2}.
\end{eqnarray}
The minimal eigenvalue $\mu_c^2$ is obtained by minimizing the expression
\begin{equation}\label{SLeigenvalue}
\mu_c^2=\frac{\int_0^1 (\mathcal{T} {F'}^2-\mathcal{P}F^2)dz}{\int^1_0 \mathcal{Q} F^2 dz}
\end{equation}
with the boundary condition $F'(0)=0$~\cite{Cai:2011ky,Li:2013fza}.
Considering comprehensively the boundary conditions $F(0)=1$ and $F'(0)=0$, we introduce a trial function with the same form to (\ref{trialfunction}) and thus read off the critical chemical potential from Eq.~(\ref{SLeigenvalue}).
Concretely, we plot the critical chemical potential $\mu_c$ versus the CAC parameter $\alpha$ in the form of solid points in the left plot of Fig.~\ref{Criticalmugrandpotential} and list the results in Tab.~\ref{tab:ssTcna}.
\begin{table}
\caption{\label{tab:ssTcna} The analytical results for the insulator/superconductor model with $\Delta=2$: the critical chemical potential from the analytical method ($\mu_c$), $\langle \mathcal{O} \rangle \approx \mathcal{C}_{3a}\sqrt{\frac{\mu-\mu_c}{\mu_c}}$ and $\rho\approx \mathcal{C}_{4a}(\mu-\mu_c)$ near the critical point. }
\begin{ruledtabular}
\begin{tabular}{c c c c c c c  }
  $\alpha$ & 0.01 &  0.2 &0.5 & 1&5&10  \\ \hline
  $\mu_{ca}$&1.7080&$1.5582$&1.4241&1.2955&$0.9690$&$0.8374$\\
   $\mathcal{C}_{3a}(\mu\approx \mu_c)$&1.4247&$1.3725$&1.2954&1.2053&$0.9345$&$0.8148$\\
   $\mathcal{C}_{4a}(\mu\approx \mu_c)$&1.0604&$1.1364$&1.1826&1.2157&$1.2701$&$1.2841$\\
 \end{tabular}
\end{ruledtabular}
\end{table}
It is observed that the analytical results agree well with the numerical ones, which means that the S-L method is still powerful in the CAC insulator/superconductor model.

When the chemical potential is slightly above the critical point, the condensate $\langle \mathcal{O}\rangle$ is very small, so we can expand the gauge field $\phi(z)$ in the form of the small parameter $\langle \mathcal{O}\rangle$ as
\begin{equation}\label{solitonphiochi}
    \phi(z)=\mu_c+\langle \mathcal{O}\rangle \chi(z)+\cdots.
\end{equation}
Combining with Eq.~(\ref{ConFz}), we have  the equation of $\chi(z)$ as
\begin{equation}\label{schizprim2}
\chi''-\left(\frac{2}{z}+\frac{f'(z)}{f(z)}\right)\chi'-\frac{2  \mu _c \langle \mathcal{O}\rangle \left(1-\beta z^2\right)^2 z^{2 \Delta -4}}{f(z)}=0.
\end{equation}
Defining  $T(z)= z^2 f(z)$, Eq. (\ref{schizprim2}) can be rewritten as
\begin{equation}\label{tchieqanasol}
(T\chi')'=2  \mu _c \langle \mathcal{O}\rangle \left(1-\beta z^2\right)^2 z^{2 \Delta -2}.
\end{equation}

At the boundary~($z\rightarrow0$), the function $\chi(z)$ can be further expanded as
\begin{equation}\label{chiexpand}
\chi(z)=\chi(0)+\chi^\prime(0)z+\frac{1}{2}\chi^{\prime\prime}(0) z^2+\frac{1}{6}\chi^{\prime\prime\prime}(0)z^3+\cdots.
\end{equation}
Substituting Eq.~(\ref{chiexpand}) into Eq.~(\ref{solitonphiochi}) and thus comparing it with Eq.~(\ref{asySwaveBHphi}), we can get
\begin{eqnarray}
  \langle\mathcal{O}\rangle &=& \frac{1}{\chi(0)}(\mu-\mu_c), \label{anaOchi} \\
  \rho &=&-\chi'(0) \langle\mathcal{O}\rangle=-\frac{\chi'(0)}{\chi(0)}(\mu-\mu_c).\label{anarhochi}
\end{eqnarray}
Obviously, the following important thing is to calculate the values of $\chi'(0)$ and $\chi(0)$.
By using the boundary condition $\chi'(1)=0$~\cite{Cai:2011ky,Li:2013fza}, integrating Eq.~(\ref{tchieqanasol}) reads
\begin{equation}
(T\chi')|^z_1= \mu _c \langle\mathcal{O}\rangle\int^{u=z}_{u=1}2\left(1-\beta u^2\right)^2 u^{2 \Delta -2}du=\mu _c \langle\mathcal{O}\rangle \mathcal{C}_7(\beta,\Delta, z).
\end{equation}
Further integrating the above equation, we can obtain the value of $\chi(0)$ as
\begin{equation}\label{chi0anas}
\chi(0)=\mu _c \langle\mathcal{O}\rangle \int^0_1\frac{\mathcal{C}_7(\beta,\Delta, z)}{T}dz=\mu _c \langle\mathcal{O}\rangle\mathcal{C}_8(\beta,\Delta).
\end{equation}
Meanwhile, taking the limit $z\rightarrow0$ yields the value of $\chi'(0)$ as
\begin{equation}\label{chiprime0anas}
\chi'(0)=\mu _c \langle\mathcal{O}\rangle \lim_{z\rightarrow0}\frac{\mathcal{C}_7(\beta,\Delta, z)}{T}=\mu _c \langle\mathcal{O}\rangle\mathcal{C}_9(\beta,\Delta).
\end{equation}
Combining Eqs.~(\ref{chiprime0anas}) and (\ref{chi0anas}) with Eqs.~(\ref{anaOchi}) and (\ref{anarhochi}), the scalar condensate and the charge density near the critical chemical potential can be expressed as
\begin{eqnarray}
\langle\mathcal{O}\rangle &=&\frac{1}{\sqrt{\mathcal{C}_8}}\sqrt{\frac{\mu}{\mu_c}-1}=\mathcal{C}_{3a}\sqrt{\frac{\mu}{\mu_c}-1},\label{anaOfinal} \\
\rho &=&-\frac{\chi'(0)}{\chi(0)}(\mu-\mu_c)=-\frac{ \mathcal{C}_9(\beta,\Delta)}{\mathcal{C}_8(\beta,\Delta)}(\mu-\mu_c)=\mathcal{C}_{4a}(\mu-\mu_c).\label{anarhofinal}
\end{eqnarray}
We calculate and list the values of $\mathcal{C}_{3a}$ for Eq.~(\ref{anaOfinal}) and $\mathcal{C}_{4a}$ for Eq.~(\ref{anarhofinal}) in Tab.~\ref{tab:ssTcna} to compare with the numerical results in Tab.~\ref{criticalmuc1c2}.
It is obvious that the analytical results agree with the numerical ones at the same order, especially, the trend of the  effect for CAC parameter on the coefficient is consistent with each other. What is more, the  results in the case of $\alpha=0.01$  almost recover the ones in Refs.~\cite{Nishioka131,Lu:2013tza,Cai:2011ky,Li:2013fza}.

\section{Conclusions and discussions}
In the probe limit, we have constructed the holographic $s$-wave superconductor models in the four-dimensional CAC black hole and soliton backgrounds via both numerical and analytical methods. The effects of the CAC parameter $\alpha$ on the superconductor models were studied in detail and the main conclusions  are as follows.

In term of the conductor/superconducotor model, when the scaling dimension is fixed as $\Delta=2$, the critical temperature increases with the larger CAC parameter $\alpha$, which means that the increasing CAC parameter enhances the superconductor phase transition. Meanwhile, the critical exponent of the condensate is $\frac{1}{2}$, which suggests that the system suffers from a second-order phase transition at the critical point, and thus be upheld by the behavior of the grand potential.
What is more, below the critical point, an obvious pole appears in the low frequency region for the imaginal part of conductivity, which corresponds to a delta function of the real part of the conductivity and thus implies the infinite DC conductivity expected from the  superconductor. Furthermore, from the minimum of the imaginal part of conductivity, we read off the energy gap, which decreases with the increasing CAC parameter and is consistent with the behavior of the condensate. In addition, the analytical results such as the critical temperature, the critical exponents of condensate $\langle \mathcal{O}\rangle$ agree with the numerical results, and the coefficients of  $\langle \mathcal{O}\rangle$ is qualitatively the same with the numerical ones near the critical point.

As for the insulator/superconducotor model, the critical chemical potential with $\Delta=2$ decreases with the increasing CAC parameter $\alpha$, which means that the increasing CAC parameter enhances the superconductor phase transition. Meanwhile, the critical exponent of the condensate ($\frac{1}{2}$)  suggests that a second-order phase transition occurs at the critical point, which is testified by  the grand potential. What is more, near the critical point, the charge density increases linearly with the chemical potential, which is the university of holographic insulator/superconductor model. Furthermore, beyond the critical point, the imaginal part of conductivity displays an obvious pole at the low frequency region, which implies that the system is indeed at superconducting state above the critical point. In addition, the hairy state is proved to be stable compared with the no-hair state from the analysis of the grand potential. The analytical results such as the critical chemical potential, the critical exponents of condensate $\langle \mathcal{O}\rangle$ and charge density $\rho$ agree with the numerical results, and the coefficients of  $\langle \mathcal{O}\rangle$ and  $\rho$ are qualitatively the same with the numerical ones near the critical point.
Even though  the present calculation are restricted to some special cases of the parameter space of scaling dimension $\Delta$ and CAC parameter $\alpha$,  we can obtain the qualitatively same results for other values of $\Delta$ and $\alpha$.

Comprehensively speaking, the increasing CAC parameter enhances both the conductor/superconductor phase transition and the insulator/superconductor  phase transition for fixed scaling dimension. What is more, for both models, there always exists a critical value. Near the critical point,  the system suffers from a second-order phase transition expected from the mean-field theory. Meanwhile, the state with scalar condensate is confirmed to be thermodynamically stable. Furthermore, all above numerical results are backed up by the analytical results.
In addition, as discussed in Sec.~I, the present work also investigated the effects of curvature correction on holographic superconductors in the four-dimensional Gauss-Bonnet gravity in some sense. However, it should be noted that the increasing curvature correction parameter $\tilde{\alpha}$ always hinders the $s$-wave conductor/superconductor phase transition in the range $\tilde{\alpha}<0$~\cite{Qiao:2020hkx}, while
the increasing CAC parameter enhances the superconductor phase transition in the current paper. 
By analysing the metric functions of the CAC gravity and the four-dimensional Gauss-Bonnet one~\cite{Qiao:2020hkx}, this inconsistency is reasonable,
because the CAC parameter $\alpha$ equates to minus one half of the Gauss-Bonnet parameter $\tilde{\alpha}$ in Ref.~\cite{Qiao:2020hkx}, i.e, if we define $-2\alpha=\tilde{\alpha}$, the current metric function~(\ref{CACBH}) restores to the Gauss-Bonnet metric function~(6) in Ref.~\cite{Qiao:2020hkx}. Of course, after this definition, we can easily forecast that the increasing CAC parameter does not enhance the phase transition but hinders the condensate to appear.

\acknowledgments
We would like to thank Prof.~L.~Li for his helpful discussions and comments. This work is supported in part by NSFC (Nos.~11865012, 11647167, 11575075 and 11747615), Foundation of Guizhou Educational Committee(Nos. Qianjiaohe KY Zi [2016]311 Zi),  Foundation of Scientific Innovative Research Team of Education Department of Guizhou Province (QNYSKYTD2018002), Program for the Natural Science Foundation of Shanxi Province, China(Grant No.201901D111315) and the Natural Science Foundation for Young Scientists of Shanxi Province,China (Grant No.201901D211441)..

\end{document}